\documentclass[twocolumn]{aa}			
\usepackage{graphicx}

\usepackage{txfonts}
\usepackage{natbib}
\bibpunct{(}{)}{;}{a}{}{,}
\newcommand{\msun}{$\rm M_{\odot}$}
\newcommand{\Lsun}{$\rm L_{\odot}$}
\newcommand{\mhi}{$M_{\rm HI}$}
\newcommand{\lb}{$L_{\rm B}$}
\newcommand{\hi}{H\,{\sc i} }

\begin{document}
   \title{The \hi content of the
  advanced merger NGC\,4441}


   \author{E. Manthey
          \inst{1,2}
          \and
          S. Aalto\inst{3}
          \and
          S. H\"uttemeister
          \inst{1}
          \and
          T.~A. Oosterloo
	  \inst{2,4}
          }

   \offprints{manthey@astron.nl}
  \institute{ 
University of Bochum, Department of Astronomy, 44780 Bochum, Germany\\
             \and 
Astron, 7990AA Dwingeloo, The Netherlands\\
         \and
             Onsala Space Observatory, Chalmers University of Technology, SE-439
92 Onsala, Sweden\\
\and
Kapteyn Astronomical Institute, University of Groningen, P.O. Box 800, 9700AV Groningen, The Netherlands
             }

   \date{Received ; accepted }

 
  \abstract
   {NGC\,4441 is a candidate for a merger between a spiral and an
   elliptical galaxy (S+E merger), because it shows typical tidal structures
   such as an optical tail and 
   two shells. With a far-infrared luminosity of
$\rm \sim5\cdot10^9\,L_{\odot}$ this galaxy belongs to the class of
moderate luminosity mergers, in which the merging process
induces (if at all) only a moderate starburst.}
{The study of the atomic gas content allows us to investigate 
the merger history and the impact on the star formation. In
   particular, it is not clear from simulations whether an S+E
 merger leads to a
   gas 
   concentration in the nucleus, resulting in a starburst, or whether the gas
 is
   spread out and therefore too diffuse for new star forming regions.}
 {We used the Westerbork
Radio Synthesis 
Telescope to observe the properties of the H\,{\sc i}. 
By using this interferometer,
we are able to study the large-scale \hi distribution and kinematics with
high spatial and velocity resolution.  }
{We found two \hi tails extending out to more than 40\,kpc. In a 
central disc, the gas shows a fairly regular rotation pattern indicating that the
 gas started to settle after the merger. The 
total \hi mass adds up to $\rm
1.5\cdot10^9\,M_{\odot}$. By comparing
   the high resolution \hi maps with deep optical images, differences between
   the stellar and gaseous tidal features are apparent, which
indicates an S+E merger origin.
}
   {}

   \maketitle
%

\section{Introduction}
 Extreme
mergers between two large disc galaxies, leading to a super--starburst and to
ULIRGs \citep{1996ARA&A..34..749S} have been at the focus of research, but
those 
galaxies are rare.  Small spirals and ellipticals dominate the galaxy
population in many environments, but mergers between them, so-called S+E
mergers, have been poorly 
studied so far. 
Models of S+E mergers are
in strong disagreement 
concerning the prediction of enhanced star formation induced by the merger.
\citet{1993ApJ...405..142W} used a smoothed particle hydrodynamics approach
 (SPH) to model the gas. With this fluid description, the stellar and gaseous
components are quickly
segregated. While the stars form features like shells, the gas congregates in
the centre of the 
remnant galaxy, leading to a strong gas concentration and thus resulting in a
starburst, similar to ULIRGs. 
Simulations by \citet{1997ApJ...481..132K} however predict a dispersion of the gas clouds
which might 
not lead to a starburst at all, because the density of the gas is too low to
collapse and form new stars. They model the gas component as a system of
inelastic cloud particles dissipating kinetic energy in mutual collisions.
If the timescale for cloud-cloud collisions is relatively long, the cloud system
behaves like a collisionless system, thus like the stellar component.
Therefore, the gas and stellar distribution are similar. 
 While the contrast of the structures is somewhat  higher for the
stellar component, both show shell--like features, as were also found for the
stars 
in the models of \citet{1993ApJ...405..142W}. 
\\
Simulations by \cite{2000ASPC..197..267N}, \cite{2001ASPC..230..453N}, 
\cite{2003ApJ...597L.117K} and \cite{2006MNRAS.372..839N} revealed the 
importance of S+E mergers for the formation of bright ellipticals. As 
\cite{2003ApJ...597L.117K} claim, these aspects are, however, poorly studied 
so far, and detailed observational investigations are needed besides more
numerical simulations to better understand the process of those mergers.
\\
One 'prototypical' S+E merger candidate is NGC\,4194, the Medusa. In the
optical we see a diffuse tail going to the north and on the opposite side 2
shells are visible, as predicted by models. Molecular
gas is found in a continuous distribution out to 4.7\,kpc from the centre,
i.e much more spread out than 
in the case of a ULIRG
\citep{2000A&A...362...42A,2001A&A...372L..29A}. However, this galaxy is
clearly undergoing an intense 
starburst phase, albeit not as strong as in ULIRGs
\citep{2004AJ....127.1360W}. 
\\
\noindent
Here we present high resolution interferometric \hi data of another
S+E merger candidate, NGC\,4441. 
This galaxy is morphologically very similar to the Medusa. It possesses one
optical tidal tail and two bluish shells on the opposite side, which are, however,
brighter, 
i.e. more evolved, than the shells in the Medusa. The main body has an
elliptical shape \citep{1981A&A....97..302B} with a small dust layer through
the centre on the minor axis. 
Based on spectroscopic data \citet{1981A&A....97..302B}
claims that this galaxy is a merger where most
of the gas has been already used for star formation, which means only little
ongoing 
star formation. However, there must have been a period of enhanced star
formation, 
since the stellar population is younger than that of a normal
elliptical galaxy \citep{1981A&A....97..302B}. 
Our own optical spectra confirm this and we
estimate that a moderate starburst  occured $\sim$ 1\,Gyr ago 
\citep{2005AIPC..783..343M}. 
\\
The
  observations presented here were obtained with the Westerbork Synthesis Radio
Telescope.  We  
 refer to the analysis of optical long-slit spectra for metalicity
estimations, but these data are
presented in detail in \cite{optsample}.
\\
In Table\,\ref{n4441chap_intro} general information on NGC\,4441 is given.
In Section 2 we give basic information about the observations obtained, in
Section 3 we describe the data reduction for the interferometric data. 
Results are presented in Section 4, including derived
properties like the \hi gas mass and the star formation rate based
 on
the 
20\,cm continuum flux. In Section 5 we discuss the observations with particular
emphasis on models of mergers between Ellipticals and Spirals and we compare
NGC\,4441 
with the S+E merger prototype NGC\,4194 (the Medusa). Finally, in Section 6 we 
summarise the
observations and results.

\begin{table}
      \caption[Basic properties of NGC\,4441]{Basic properties of
        NGC\,4441. The distance is based on ${\rm 
        H_0 = 75\,km\,s^{-1}\,Mpc^{-1}}$.
        }
        \label{n4441chap_intro}
\centering
\begin{tabular}{lc}
\hline\hline
 property& \\
\hline
RA (2000) & 12:27:20.3\\
DEC (2000) & +64:48:05\\
$v_{\rm opt,hel}$ (km\ s$^{-1}$) &  2722\\
$D$ (Mpc) & 36 \\  
type & SAB0+ pec \\
$L_{\rm B} \ (\rm 10^{9}\,L_{\odot})$ &10.1 \\
$L_{\rm FIR} \ (\rm 10^9\,L_{\odot})$ & 5.4 \\
1\arcmin & 10.5\,kpc \\    
\hline
\end{tabular}
\end{table}


\section{Observations}

\subsection{WSRT}
The \hi observations with the Westerbork Synthesis Radio Telescope
 (WSRT) were  carried out in
 February 2003. 
 Table\,\ref{n4441chap_wsrtobs}
summarises the most important observing parameters. We observed NGC\,4441 in
one 12\,h run in the maxi--short array configuration. This is best suitable
 for observations of an extended source within a single 12 hours track. The
 quality of the data is 
good, no interference problems occurred. For bandpass and flux calibration,
 the calibrator 3C48 was observed before the observations.
Note that no secondary phase calibrator was observed. Instead, phase
calibration at the WSRT is usually done using self-calibration.
The chosen setup with a total bandwidth of 20\,MHz (corresponding to
 $\sim$4000\,km\ s$^{-1}$) and a velocity resolution 
 of 4\,km\ s$^{-1}$channel$^{-1}$ was capable to detect both narrow features in velocity
 space and a broad total velocity range due to tidal disruption. With the
 chosen setup, two polarisations were observed, with 1024 channels available
 for each of them.

\begin{table}
\caption[WSRT \hi observations of NGC\,4441]{Parameters for the WSRT
  observations.} 
\label{n4441chap_wsrtobs}
\centering
\begin{tabular}{lc}
\hline\hline
 observing parameters  &     \\
\hline
date &28.2.2003  \\
centre frequency& 1408.9  \\
total bandwidth& 20\,MHz \\ 
number of channels& 1024 \\  
velocity resolution (km\ s$^{-1}$)& 4.12 \\   
rms noise level (robust=0.5) (mJy\ beam$^{-1}$) & 0.4\\
beamwidth (\arcsec)& 20 x 19\\
primary beam & $\rm 0.6^{\circ}$ \\
primary calibrator&  3C48\\
\hline
\end{tabular}
\end{table}

\subsection{Optical imaging}
We present here an optical R band image which was taken with CAFOS at the
2.2\,m telescope on Calar Alto, Spain (CAHA) in April
2004.  
To study very faint morphological features, we obtained deep imaging with
an integration time of 60 minutes. Standard reduction using IRAF was applied
to the data. Here we use the optical image to compare the stellar and gaseous
distribution. For further analysis of the optical data, we refer to
\citet{optsample}. 

\section{Data Reduction}
The \hi data were reduced using
 MIRIAD\footnote{http://www.atnf.csiro.au/computing/software/miriad/} with
 some additional tasks 
 written by T. Oosterloo which are specific to the Westerbork
 interferometer\footnote{http://www.astron.nl/$\sim$oosterlo/wsrtMiriad/}. 
 After inspecting and, where necessary, flagging the
calibrators, 
the WSRT data have to be calibrated first for changes in the system 
temperature. Since no
secondary calibrator was observed, after flux and bandpass calibration a
continuum image was created to perform self-calibration on continuum sources
within the field.
 To achieve a proper calibration, it is necessary to deeply
CLEAN the continuum image. In an iterative process, first a mask was created 
which included only real
emission to define the CLEANing components. Second, the image was CLEANed
using the mask and subsequently a new, deeper mask could be generated for a
better CLEANing. Once the image was finally CLEANed, self-calibration was
performed, and the loop started again with creating a mask, CLEANing and
self-calibration. The data quality was high enough that only 3 iterations
were necessary for a proper calibration. The results were then applied to the
target galaxy.
 After 
continuum
subtraction, a datacube was created to which we applied Hanning smoothing,
leading to a velocity resolution of 8\,km\ s$^{-1}$.
 The datacube was convolved with a Gaussian of roughly 
double the
size of the original beam. With this smoothed image a mask was created
including real emission only. This was checked for all channels using the
KARMA movie option \citep{1996ASPC..101...80G}. This mask was used for 
CLEANing as described above for the self-calibration process of the continuum. This
CLEANing procedure was repeated 3 times until no significant sidelobes were
 visible.  Integrated intensity and velocity maps were created using the MOMENT
task in MIRIAD. Finally, primary beam correction was applied to the
 integrated intensity 
maps. We created maps which were naturally weighted, thus providing the
 highest sensitivity, as well as maps with a robust weighting of 0.5 to reach
 a higher spatial resolution.  We reached an rms of 0.4\,mJy\ beam$^{-1}$ in the
 naturally weighted map and and rms of 0.5\,mJy\ beam$^{-1}$ in the robustly weighted
 map. Here we present only the robust map with the higher spatial resolution,
 because the naturally weighted maps, although more sensitive, do not show any
additional significant flux contribution compared to the robust map.

\section{Results}
\subsection{Neutral hydrogen}
\subsubsection{Morphology of the tidal tails}

 \begin{figure*}
  \centering 
 \includegraphics[angle=0,width=5.9cm]{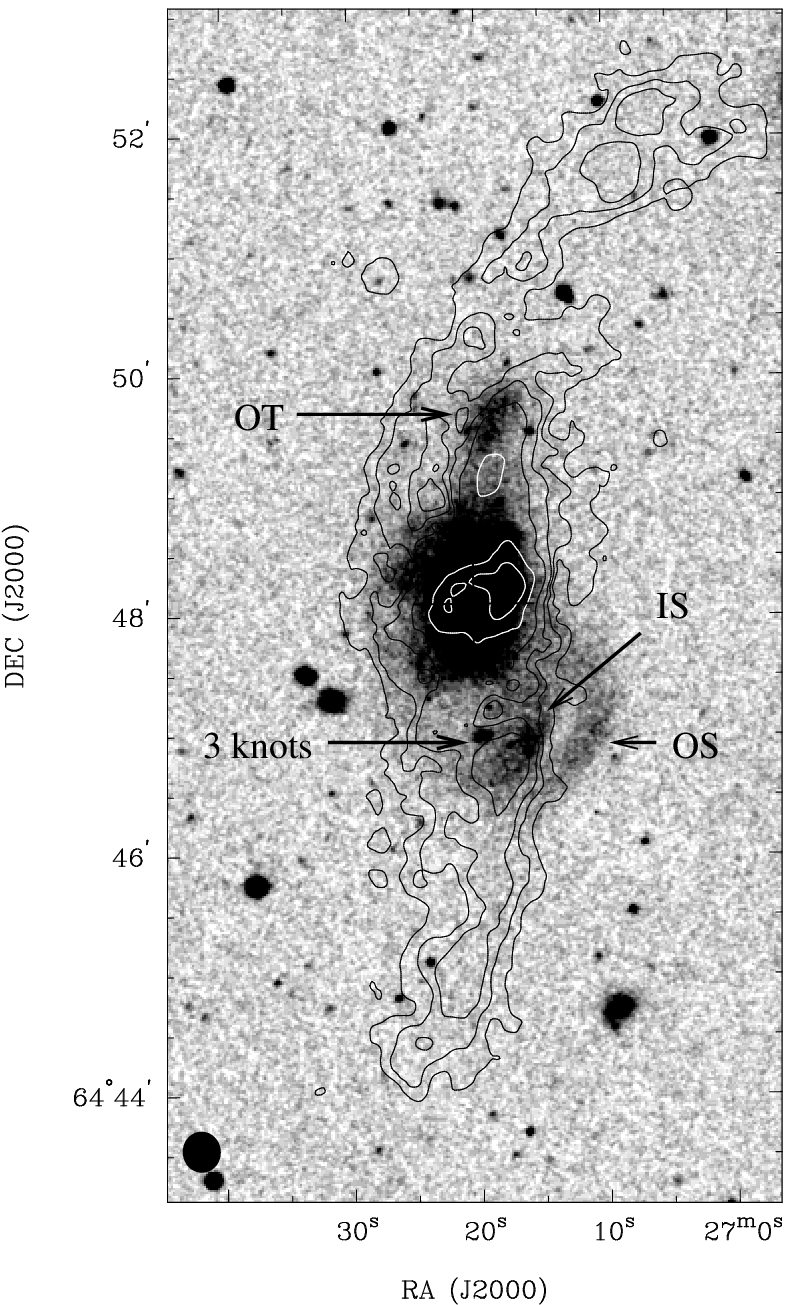}
  \includegraphics[angle=0,width=7cm]{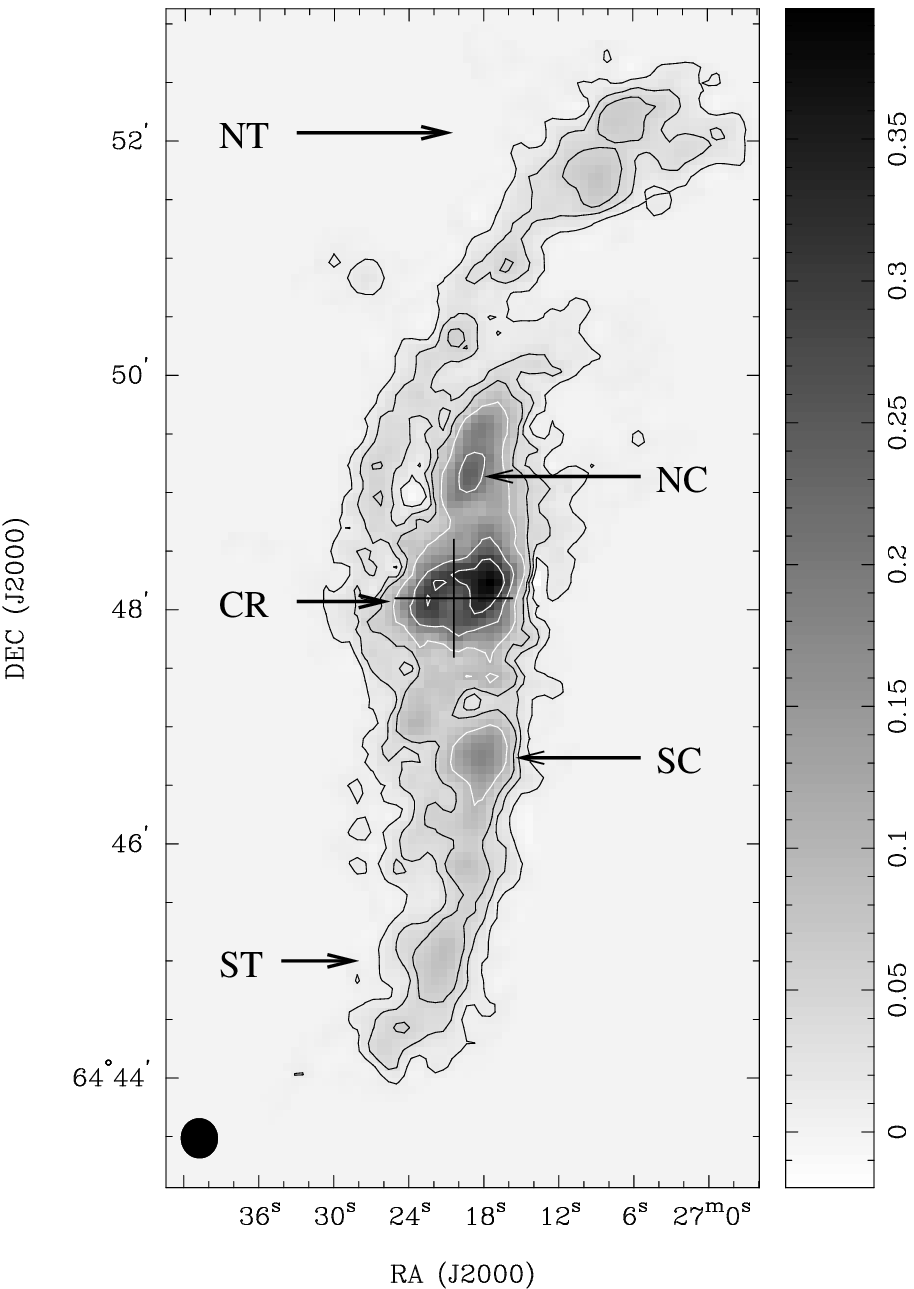}
   \includegraphics[angle=0,width=7cm]{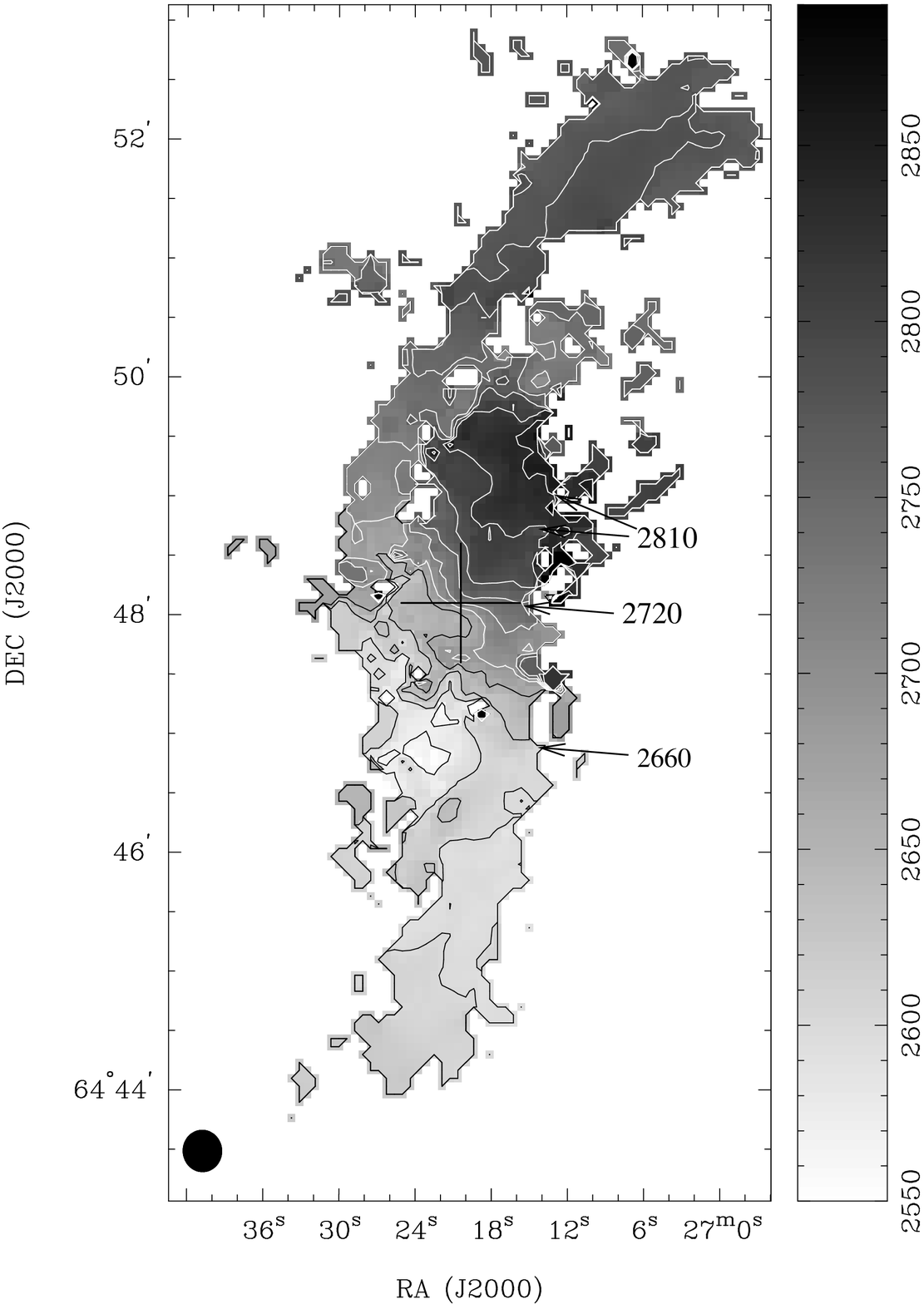}
   \includegraphics[angle=0,width=7cm]{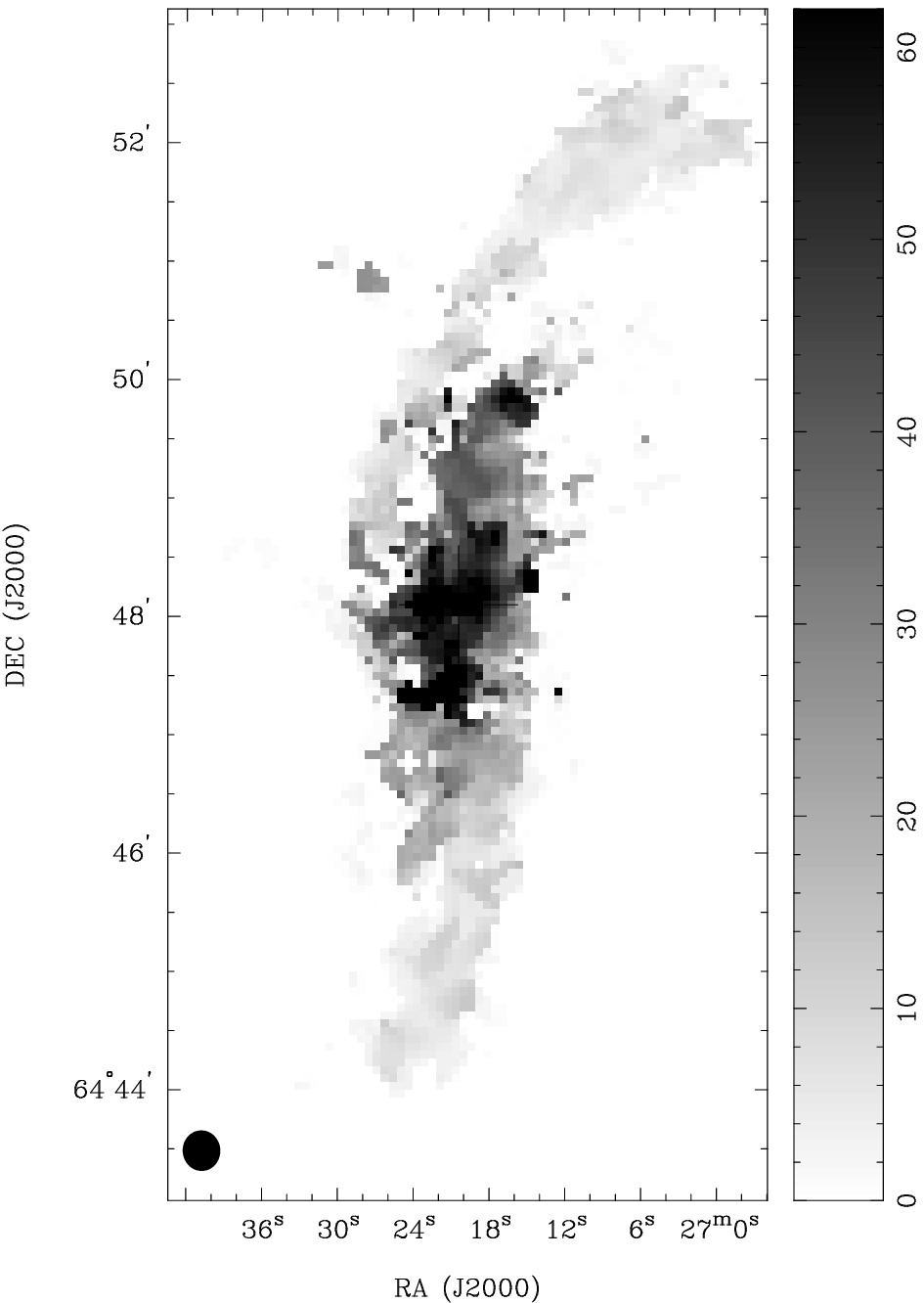}

      \caption[NGC\,4441: \hi data]{NGC\,4441 -- {\bf top left:} 
              \hi distribution overlayed on optical DSS image. 
              Contour levels are 
              $\rm 2.8, 8.4, 14, 28, 56, 84 \times 10^{19}\,cm^{-2}$
Marked
are the optical features: the optical tail (OT), the inner shell (IS), the
outer shell (OS) and the three optical knots, {\bf top right}
 \hi distribution, contours same as above. Marked are the gaseous
features as described in the text: the northern tail (NT), the northern clump
 (NC), the central ring (CR), the southern clump (SC), and the southern tail 
(ST), {\bf bottom left:}
    velocity field, contour levels are 
2570 to 2840\,km\ s$^{-1}$, in steps of 30\,km\ s$^{-1}$,
              {\bf bottom right:}
              \hi velocity dispersion (2nd moment).
              The cross marks the optical centre position. 
              }
         \label{n4441chap_n4441hi}
   \end{figure*}

 \begin{figure*}[h!]
   \centering
   \includegraphics[angle=-90,width=16cm]{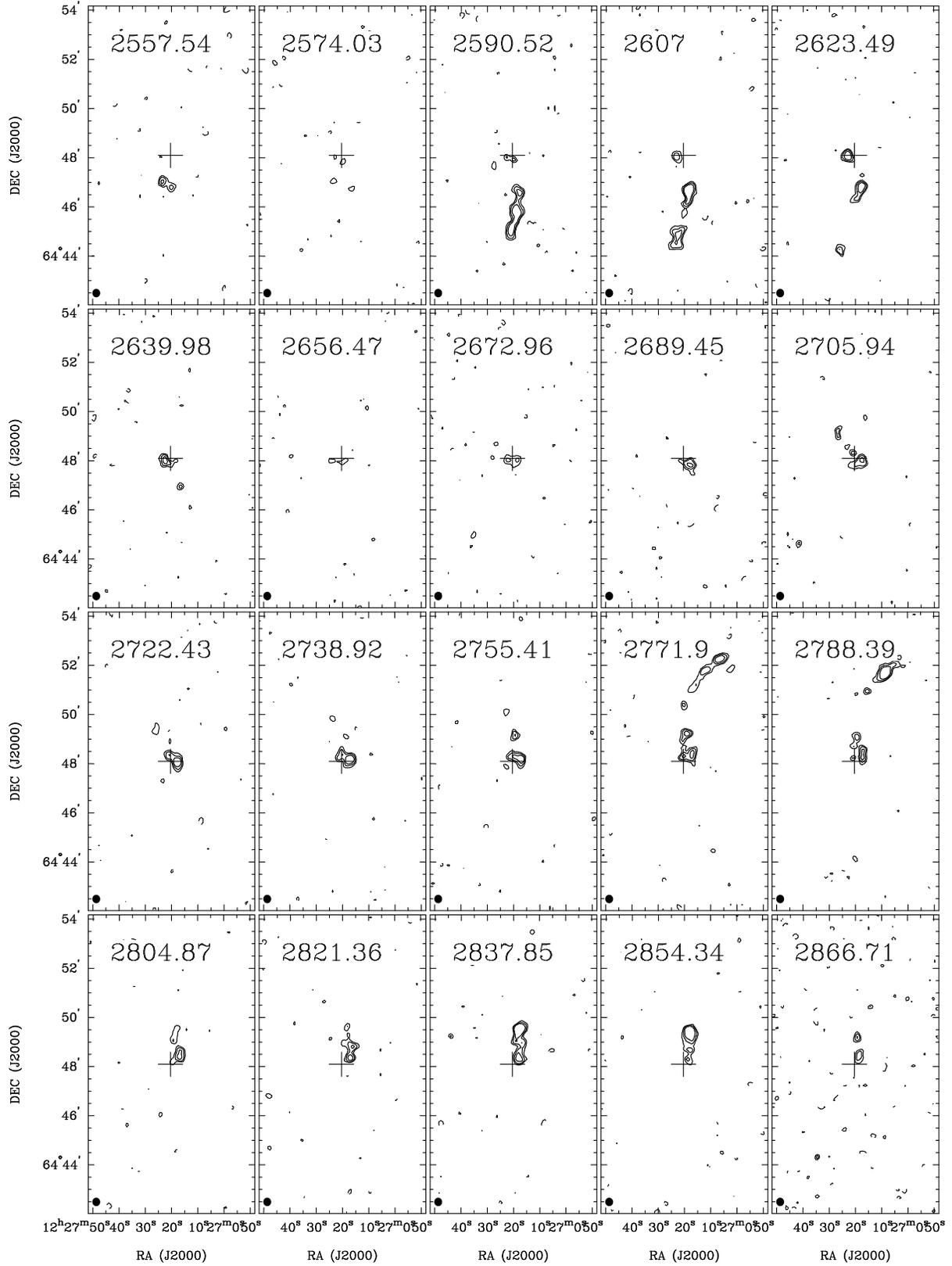}
 \caption[NGC\,4441: \hi channel maps]{Channel maps (robust weighted) of
            NGC\,4441. The 
            optical centre 
            position is marked with a cross. The contour levels are
            -0.9,1.1, 1.3, 1.5, 1.7, 1.8, 2\,mJy\ beam$^{-1}$.
 The heliocentric velocity is
            given at the top 
            left, the beam  at the bottom left corner.
            }
         \label{n4441chap_n4441chanmap}
   \end{figure*}

\begin{table}
\caption[\hi properties of NGC\,4441]{The \hi properties of the total
system, the tails and the central disc of NGC\,4441. Heliocentric
velocities are given.\\ 
 }
\label{n4441chap_himass}
\centering
\begin{tabular}{lc}
\hline\hline
 \hi properties & \\
\hline
{\bf total system} & \\
 ${\rm v_{opt,hel}}$ (km\ s$^{-1}$)& 2674\\
 vel. range (km\ s$^{-1}$) & 2580 -- 2900\\  
 $ F_{\rm HI}$ (Jy\,km/s) & 4.83\\
\mhi ($10^9$\,\msun) & 1.46\\  
\mhi / \lb & 0.82\\
\hline
{\bf central ring} & \\
major axis (\arcsec, kpc) & 32, 5.5\\
minor axis (\arcsec, kpc) & 19, 3.4\\
\mhi ($10^8$\,\msun) & 4.2\\
\hline
{\bf northern tail} & \\
vel. range (km\ s$^{-1}$) & 2700 -- 2820\\
extent (centre--tip) (\arcmin, kpc)& 4.6, 48\\
centre -- clump (\arcmin, kpc) & 1.5, 16 \\
 \mhi ($10^8$\,\msun) & 2.1\\
\hline
{\bf southern tail}\\
vel. range (km\ s$^{-1}$) & 2580 -- 2640\\
extent (centre--tip) (\arcmin, kpc)& 4.0, 42\\
centre -- clump (\arcmin, kpc) & 1.7, 18 \\
\mhi ($10^8$\,\msun)& 3.4\\
\hline
\end{tabular}
\end{table}

The neutral hydrogen distribution in NGC\,4441 shows two prominent tidal tails
to the north and to the south (see Fig.\,\ref{n4441chap_n4441hi}). The southern 
arm is
blueshifted with respect to the systemic velocity
(see Fig.\ref{n4441chap_n4441chanmap}), while the
northern arm has a higher velocity. 
Both tails have a similar extent:
the northern tail out to 48\,kpc, the southern tail out
to 42\,kpc
 from the nucleus and both form a relatively symmetric
structure. The \hi 
tails are clumpy, including the tip of both tails. Two large 
clumps are found, both $\sim$ 17\,kpc from the centre (marked as NC and SC 
in Fig.\,\ref{n4441chap_n4441hi}). 
From the kinematics there are no indications that these clumps are or will be
separate gravitationally bound entities like those simulated
by \cite{2007arXiv0710.3867D}.
\\
It is noticeable that the symmetry in the \hi distribution is not seen in
 the optical morphology.
Here, there is a tail to the north
which coincides nicely in position and extension with the large northern \hi
clump. In the southern 
direction, however, we see a misalignment of the \hi tail and the
optical tidal features. Two shells are visible in the optical southwest of the
main body. There is
almost no overlap  of the more western, outer shell with the \hi tail. The tip
of the inner shell corresponds to the southern  \hi clump. Three optical knots
are found at that position, of which at least
two do not look point-like, i.e., are probably not foreground
stars. They 
might be star forming knots, probably too close to the galaxy to survive as
tidal dwarf galaxies, but can also be background galaxies, because our 
deep R-band 
image shows a background galaxy cluster. However, no \hi is detected
at higher velocities within our band, which may of course miss distant
galaxies. Unfortunately, the distance of the background cluster is not known. 
\\
The gas in the tails shows only little dispersion in general ($\rm <
15\,km\ s^{-1}$). The dispersion increases slightly in regions of gradients in the
atomic gas distribution, e.g., around clumps.
\\

\subsubsection{The central ring}
\begin{figure}
 \centering
\includegraphics[angle=0,width=9cm]{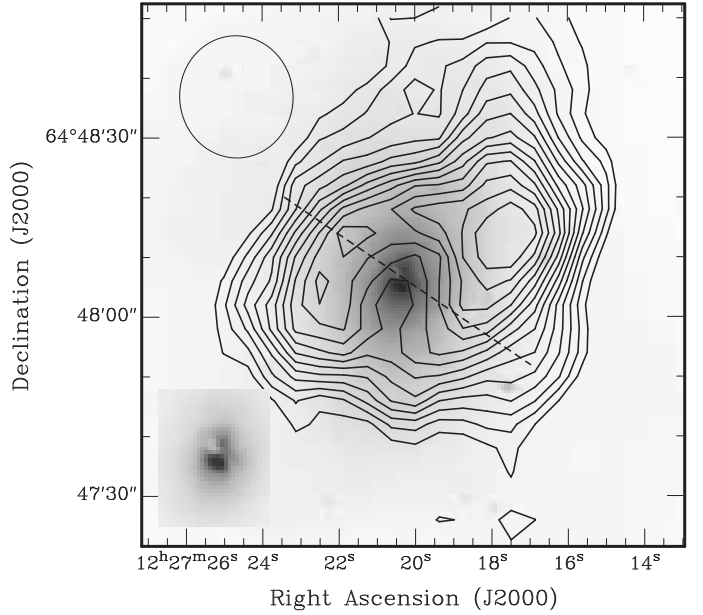}
\caption[NGC\,4441: \hi distribution in the centre]{R-band image (taken
  with the CAHA 2.2\,m telescope) of
  the central region. \hi is overlayed as 
  contours, contour levels are from 
2.8 to $\rm 10.08 \times 10^{20}\,cm^{-2}$ in steps of $\rm 
0.56 \times 10^{20}\,cm^{-2}$.
 The beamsize is shown in the upper
  left corner. The gas ring is centred on the optical nucleus, which exhibits
  a thin dust lane. The orientation of the dust lane is marked by the dashed
  line.  In the lower left corner, the
  R-band image of the nucleus with the dust lane is shown. 
}
 \label{n4441chap_ring}
\end{figure}

Fig.\ref{n4441chap_ring} shows the central region of NGC\,4441 in the 
optical R band and the \hi distribution
(robust=0.5) overlayed as contours.
We find a distinct \hi ring with a diameter of $\sim$ 11\,kpc
centred 
on the optical (and CO) nucleus. 
At the centre of the ring
there is an \hi depression, which is, however, unresolved.
 The highest
concentration of \hi is found on the west side of the ring, thus, west of the
optical centre. 
Three \hi peaks are present in the ring, with a distance of $\sim$2.5\,kpc from
the 
centre. 
This ring structure might actually be a small disc, where in the inner part most
of the gas is present in the molecular phase which is often found at
column densities of $\rm \sim 10^{21}\,cm^{-2}$ \citep{2000ASPC..218..321Y}. \\

\begin{figure}
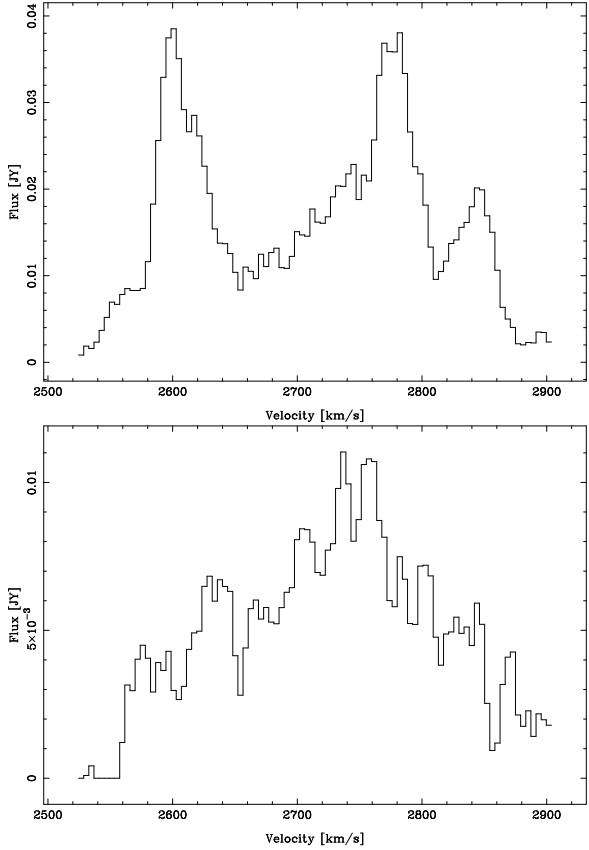

   \centering
\includegraphics[angle=-90,width=7.7cm]{7584fig07.ps}
\includegraphics[angle=-90,width=7.7cm]{7584fig08.ps}

 \caption[\hi spectra of NGC\,4441]{Integrated \hi spectrum of {\bf top:} the
            total amount of \hi in 
            NGC\,4441 and {\bf bottom:} the central ring. 
In the total \hi spectrum, the two narrow
            features  around 2600\,km\ s$^{-1}$ and 2780\,km\ s$^{-1}$ 
correspond to the two tidal tails. The third peak at 2840\,km\ s$^{-1}$ 
appears at the position of the northern clump. 
            }
         \label{n4441chap_n4441hispec}
   \end{figure}

\noindent
Fig.\,\ref{n4441chap_n4441hispec} shows the integrated \hi spectra of the 
total \hi amount and the central ring. The total \hi spectrum shows three
peaks. The features  around 2600\,km\ s$^{-1}$ and 2780\,km\ s$^{-1}$ 
correspond to the southern and northern tidal tail. A third peak at
 2840\,km\ s$^{-1}$ 
appears at the position of the northern clump (see also 
Fig.\,\ref{n4441chap_n4441chanmap}).

\subsubsection{Kinematics}
\begin{figure}[h!]
 \centering
   \includegraphics[angle=0,width=9cm]{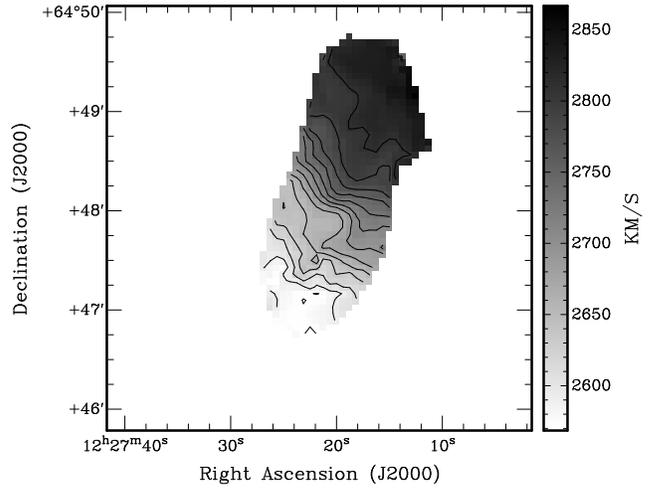}
\caption[NGC\,4441: Rotation curve]{ Inner part of 
velocity field of
 NGC\,4441, which shows a fairly regular solid body rotation
pattern. The contour levels are from 2775 to 2825\,km\ s$^{-1}$, in steps of
25\,km\ s$^{-1}$ 
}
 \label{n4441rotfield}
\end{figure}

\noindent
The main body of the \hi distribution shows a relatively regular rotation 
pattern 
which 
allows at least a crude estimate of the dynamical mass. Note that 
this region is much larger than the central ring seen in the \hi 
distribution map and has a different position angle.

\begin{table}
\caption[NGC\,4441 Rotcur results]{Estimated parameters of the central \hi disc in 
NGC\,4441.} 
\label{n4441rotcur}
\centering
\begin{tabular}{lc}
\hline\hline
{\bf estimated parameters} & \\
disc extent (\arcsec) & 60 \\
max. velocity (km\ s$^{-1}$) & 163 \\
position angle ($^{\circ}$)& -35 \\
inclination ($^{\circ}$)& 70 \\
\hline
$\rm M_{dyn}$ ($\rm 10^{10}\,M_{\odot}$)& 6.4\\ 
$M_{\rm HI}/M_{dyn}$ & 0.028\\
$M_{\rm dyn}/L_{\rm B}$ ($\rm M_{\odot}/L_{B,\odot}$) & 6.3\\
\hline
\end{tabular}
\end{table}

\begin{figure}[h!]
   \centering
\includegraphics[angle=0,width=9cm]{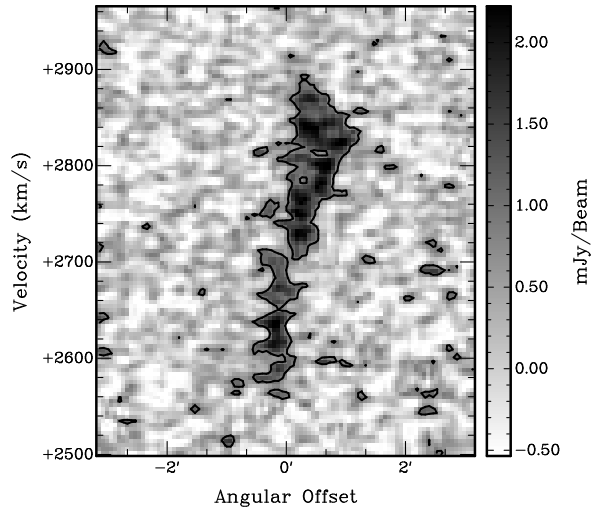}
      \caption[NGC\,4441: Position-velocity diagram]{Position-velocity
              diagram of the major axis of the inner \hi disc in
              NGC\,4441. 
The position angle is $\rm 325^{\circ}$.
Shown are the 3 $\sigma$ contours of 0.9\,mJy\ beam$^{-1}$. 
              }
         \label{n4441chap_pv}
   \end{figure}

In Fig.\,\ref{n4441rotfield} the region of the central velocity field is shown.
Using the axis ratio it is possible to estimate geometrically the inclination. Even though
drawing an ellipse is approximate, it is not arbitrary here.
Based on the elliptical shape of the velocity field shown in
 Fig.\,\ref{n4441rotfield}
 it is clear the the disc is seen relatively edge-on, thus
the uncertainty of the inclination does not affect the mass calculation 
significantly. 

We measured an axis ratio of $\sim 1/3$, equivalent to an 
inclination of $\sim$ 70\degr. An error of $\pm 20\degr$ leads to an uncertainty of a factor
 of 2
in mass and thus gives at least an indication of the mass range.  
 Note, that the 
minor axis is however not perpendicular to the major axis, which is a hint for 
non-radial motions 
like infall or outflow that could be triggered by the merger. One possible 
explanation would be an elongated gravitational potential, e.g., a bar or
spiral arms,
which forces the gas to move on ellipsoidal orbits. 
The resolution is, 
however, not sufficient to investigate this in more detail.
Furthermore, the minor axis corresponds to the thin dust lane seen in
the central region (see Fig.\,\ref{n4441chap_ring}).
\\
In Fig.\,\ref{n4441chap_pv} we present the position-velocity-diagram
(pv-diagram) through
the major axis 
of the \hi disc
(position angle (pa) = $\rm 325^{\circ}$). 
It shows a broad velocity range which is expected for a highly inclined disc.
In the centre there is a gap because of the local minimum in 
the \hi 
distribution. A second velocity component at high velocities appears in the 
inner part,
which demonstrates the complexity of the velocity distribution in this merger 
remnant.
\\
With an inclination of $\sim$ 70\degr \ and a maximum \hi rotation velocity of 
$\sim$ 
163\,km\ s$^{-1}$ 
at a distance of 1\arcmin (estimated from the pv- diagram),
 we derive a dynamical mass of 
$\rm \sim 6\cdot10^{10}\,M_{\odot}$ (see Table\,\ref{n4441rotcur}). 
Even though this is a crude estimate, the 
mass-to-light ratio is within the range of what is found in normal galaxies.\\

\subsection{20\,cm Continuum}

\begin{figure}
   \centering
   \includegraphics[angle=0,width=7cm]{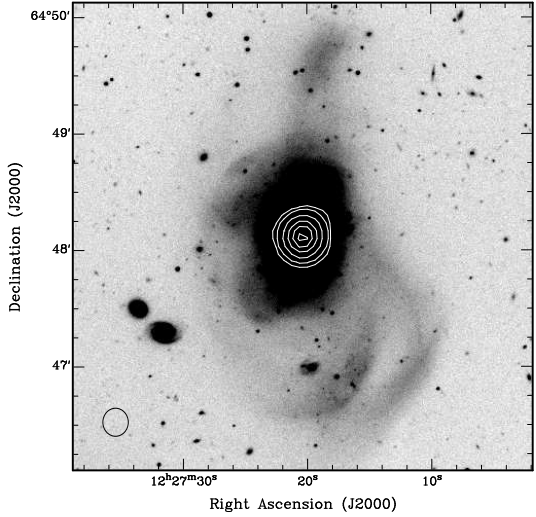}
     \caption[20\,cm continuum flux image of NGC\,4441]{R-band image (CAHA) and
              contours of the 20\,cm continuum flux. The 
              beam is shown in the lower left corner of the image. Contour
              levels are 0.5, 1, 2, 5, 7.5, 10\,mJy\,beam$^{-1}$.
              }
         \label{n4441chap_contimages}
   \end{figure}

Fig.\,\ref{n4441chap_contimages} shows an optical image with the 20\,cm radio
continuum 
overlaid as contours.
The continuum source in NGC\,4441 is 
barely resolved and concentrated on the nucleus of the galaxy. 
We fitted a Gaussian to the continuum source using the MIRIAD task IMFIT and
derived a deconvolved source size of less than 1\,kpc. The derived centre
coordinates corresponds to the optical coordinates.
To estimate the star formation rate (SFR) from the 20-cm
radio continuum flux density we use 
\begin{center}
SFR ($\rm M_{\odot}$\,yr$^{-1}$) = $0.14~D^2~F_{\rm 20\,cm}$ 
\end{center}
taken from \citet{1992ARA&A..30..575C} and \citet{2000ApJ...544..641H},
 where $D$ is the distance in Mpc and $F_{\rm 20\,cm}$ is the
20-cm radio continuum flux density in Jy. In Table\, \ref{n4441chap_contflux} the
 results for 
  NGC\,4441 are given.
Table\, \ref{n4441chap_contflux} also gives the SFR estimated from the FIR
 luminosity: 
\begin{center}
SFR ($\rm M_{\odot}$\,yr$^{-1}$) = 0.17 $L_{\rm FIR}$
\end{center}
following \citet{1998ARA&A..36..189K}, with $ L_{\rm FIR}$ in units of
$10^9$\,\Lsun. The IRAS 60\,$\mu$m and 100\,$\mu$m flux densities were taken
 from 
\citet{1990IRASF.C......0M}.
The SFR of 1--2\,$\rm M_{\odot}\ yr^{-1}$ calculated with both methods is low for a
merger and normal for 
typical spirals.
\\
 We also calculated the logarithmic ratio of FIR to
radio flux, the so-called $q$ parameter \citep{1985ApJ...298L...7H}:
\begin{center}
${q = {\rm log}\frac{F{\rm (FIR)}/(3.75\cdot10^{12}\,{\rm Hz})}{F(1.4\,{\rm GHz})}}$
\end{center}
Note, that the continuum flux at 1.4\,GHz, $F$(1.4\,GHz) in this formula, is in
  units of ${\rm W\ m^2\ Hz^{-1} =}$\linebreak
  $\rm {10^{26}\,Jy}$.
The FIR flux, $F$(FIR), is a combination of the 60\,$\mu$m and 100\,$\mu$m flux
  densities, as used for determining $L_{\rm FIR}: 
  F{\rm (FIR)} = 1.26\cdot10^{-14} \cdot (2.58 \cdot F{\rm (60\,\mu
  m)} + F{\rm (100\,\mu m)})$. 
To calculate $q$, $F$(FIR) is normalised by the mean frequency of
 $\rm 60\,\mu m$ 
  and $\rm 100\,\mu m$, which is $\rm 3.75\cdot10^{12}\,Hz$. 
\citet{1985ApJ...298L...7H} derived a mean value for $q$ of 2.3. Although
this parameter was determined for disc galaxies only (but independent of their
star formation activity), NGC\,4441 ($q=2.4$) conforms to this value.\\

\begin{table}
\caption[20\,cm continuum flux properties of NGC\,4441]{Continuum and FIR
  fluxes, SFR based on continuum and FIR emission, 
  respectively, and q parameter.}            
\label{n4441chap_contflux}      
\centering                       
\begin{tabular}{l c }       
\hline\hline            
 key characteristics & \\   
\hline                  
 20\,cm (Jy) & 0.013\\
 60\,$\mu$m (Jy) & 2.7\\
100\,$\mu$m (Jy) & 3.9\\
$SFR_{\rm 20\,cm}$ (\msun\ yr$^{-1}$) & 2.4\\
  $SFR_{\rm FIR}$ (\msun\ yr$^{-1}$) & 1.0\\
$q$  & 2.4\\
\hline                                   
\end{tabular}
\end{table}

\section{Discussion}

\subsection{The merger origin}
For an understanding of the whole merger process of galaxies resulting in 
a remnant like NGC\,4441,
 it is necessary to know the progenitors. Of course it is difficult, if not 
impossible, to get a unique answer to that question from a completely
merged remnant. However, there are some indications from our observations
which at least narrow the parameter space.\\ 
Since we found a large amount of gas in NGC\,4441, at least one spiral galaxy
must be involved in the merger process. In the following, we discuss the 
possibility of a spiral+spiral (S+S) and spiral+elliptical (S+E) merger.

\subsubsection*{Spiral+elliptical merger?}
There are several arguments that support an S+E merger origin:\\

\noindent
{\it  No strongly enhanced star formation:} Typically, S+S merger remnants
which show strong tidal features comparable to what is seen in NGC\,4441, 
host a region of a (fading) starburst, e.g., \cite{1996AJ....111..655H}. 
By looking at the FIR luminosity,
the 20\,cm continuum emission and the optical spectra 
\citep{2005AIPC..783..343M}, 
however we do not find 
any indication for a strong starburst, neither ongoing nor in the recent 
history, in NGC\,4441. Therefore, less gas is probably involved in this merger
event, which is naturally the case when one progenitor is an elliptical. \\
It has to be mentioned, however, that the strength of an induced starburst
not only depends on the available amount of gas but also on the internal
structure of the progenitor galaxies. In particular, the presence of a bulge
in a progenitor spiral galaxy will stabilise the gas disc against the formation
of a bar during the merger \citep{1999Ap&SS.266..195M}. That bar would be a 
powerful mechanism to
support infall of gas into the central region leading to a strong starburst
\citep{1994ApJ...425L..13M}.
In S+S merger simulations gas dissipates energy and thus loses angular momentum
when it becomes shocked along a bar 
\citep{1991ApJ...370L..65B,1996ApJ...471..115B,1996ApJ...464..641M,1999ASPC..182..463B}.
This leads to rapid gas inflows and a concentration of gas in the centre of 
the remnant,
resulting in a starburst. \\
 Nevertheless, the simulations 
for S+S with bulges, i.e., no merger-induced bars,
 never show such a low star formation rate as in the case
of NGC\,4441.\\
Of course, it is necessary to investigate in numerical simulations how bar 
formation
and a present bulge influences the gas in a {\it S+E merger} scenario, in 
particular
when the galaxies are completely merged. Due to the spheroidal shape of the 
elliptical
which is similar to a bulge, bar formation is likely not important and 
therefore
no strong gas inflows occur. 
Since the major and minor axes of NGC\,4441 are not perpendicular, this might indicate the
 presence of a small oval distortion which is, however, not strong enough to
trigger enhanced star formation.
\\

\noindent
{\it Theoretical models:} Various numerical simulations showed that optical
shells in combination with tidal tails are typical for S+E merger remnants
\citep{1984ApJ...279..596Q,1986A&A...166...53D,1988ApJ...331..682H,
1997ApJ...481..132K,1999ASPC..182..489C}.
These shells  are created by phase wrapping of the disc in the potential well
of the elliptical galaxy \citep{1984ApJ...279..596Q}. They occur in the 
advanced merger stage $\rm \sim 4-5
\cdot 10^8\, yr$ after the impact and last for more than 1\,Gyr 
\citep{1984ApJ...279..596Q,1997ApJ...481..132K}.\\
The simulated gas distribution strongly depends on the model used for the ISM.
 It has been noted by 
\cite{1999ASPC..182..489C} and \cite{2000ASPC..209..273C} that the often used
smoothed particle hydrodynamics (SPH) approach might 
not always be appropriate. Instead, these authors account for the clumpy 
structure of the ISM by simulating the behaviour of individual gas clouds
rather than a continuous fluid. They found a largely spread gas distribution,
often forming gaseous tails as we see in NGC\,4441. Therefore, these models
do not predict a strong starburst phase, because the gas becomes too 
dispersed. This agrees with our observations, but stands in contrast to S+S
merger scenarios \citep[e.g.,][]{1991ApJ...370L..65B,1999IAUS..186..205M}. 
\\

\noindent
{\it Stellar component similar as the Medusa:} The similarity of the optical 
morphologies of NGC\,4441 and NGC\,4194, the
prototypical S+E merger remnant, has been
mentioned in the introduction, reflecting a possible analogue merger origin.
Besides the obvious features like the stellar shells and tidal tails, both
galaxies host a dust lane.\\
In NGC\,4441, the minor axis of the fairly regularly rotating \hi disc 
corresponds to a faint
central dust lane (Fig.\,\ref{n4441chap_ring}).
It is interesting to note that also the Medusa exhibits a minor axis dust
lane. Although this may further strengthen the notion of a a similar origin,
the dust lane of the Medusa is more prominent and longer (it curves into the
optical tail). Furthermore, that dust lane is associated with a double
spectral feature \citep{2000A&A...362...42A}. Such extra features are not
seen in NGC\,4441. This may be a result of its more evolved state, because
a  large fraction of the ISM has been consumed in star formation.\\

\noindent
{\it \hi vs. optical morphology:} In NGC\,4441 we see striking differences
 between the stellar and gaseous distribution. In particular the two
optical shells stand in contrast to the two symmetric \hi tails.\\
In S+S merger remnants, however, the optical and \hi morphology are generally
similar \citep{1996AJ....111..655H}, because both the stellar and gaseous tidal
features are formed from the two discs which are merging. 
\cite{2000AJ....119.1130H} found some examples in which the optical and
the gaseous tails are displaced or where some regions are lacking gas,
 while the overall morphologies are still similar. The authors  
described several scenarios for these
discrepancies between the stellar and gaseous tidal morphology and claim that 
mainly superwinds are responsible for the
differences found. These superwinds are driven by the central starburst 
occuring in S+S merger remnants and sweep away the gas in the direction
of the wind.\\
Their findings differ, however, in two ways from NGC\,4441. First, NGC\,4441 
does not host an ongoing strong starburst, neither did it in the recent 
history. Therefore, the existence of a superwind is unlikely. Second, we see
not only an offset of the gaseous and stellar tidal features, but 
also a completely different morphology. This cannot be explained by a
possible superwind alone. Other possible explanations given by 
\cite{2000AJ....119.1130H}, like dust-obscuration and photoionisation, can only
explain local differences and not large-scale discrepancies, as pointed out by
the authors. Photoionisation might explain why there is only little gas found
in the outer optical shell. In that case we expect to find a large amount of
ionised gas which can be detected in e.g. H$\alpha$ emission. This will be done
in a further investigation.

\subsubsection*{Spiral+Spiral merger?}
There is one striking difference between NGC\,4441 and S+E merger simulations
as well as the prototype NGC\,4194:\\

\noindent
{\it Two symmetric \hi tails:} The discussion of the merger origin of N
GC\,4441
is complicated by the finding of two symmetric \hi tails. This is typical
for a merger between two disc galaxies, when each disc forms one tail during
the encounter \citep[e.g.,][]{1972ApJ...178..623T,1994mtia.conf..323H,
1996ApJ...464..641M}. Neither
 the simulations by \cite{1993ApJ...405..142W} nor 
\cite{1997ApJ...481..132K} show the appearance of two tails in an S+E merger
scenario. Furthermore, our \hi data of the prototypical S+E merger 
NGC\,4194 show only one gaseous tail in that galaxy, in accordance to the
S+E merger simulations \citep{medusa}\\
Observations of the interacting S+E galaxy system Arp\,140 
(an early stage of an S+E 
merger where both galaxies are still clearly separated) done by 
\cite{2007MNRAS.376...98C} showed as well one \hi tail only. Furthermore,
they found an \hi bridge between the interacting galaxies formed by disc
material which streams towards the elliptical galaxy 
\citep{1972ApJ...178..623T}. This bridge will likely be destroyed once the 
galaxies finally merge, as in the case of bridges found in S+S mergers
\citep{1972ApJ...178..623T,1995PhDT.........8H}.\\

\noindent
In recent deep observations of early-type galaxies \citet{2006MNRAS.371..157M}
found a significant
amount of \hi in a large
fraction of those galaxies. The data reveal a variety of structures, including
extended discs. However, the peak column density is with 
\linebreak
N(H\ {\sc i}) $\rm \leq
10^{20}\,cm^{-2}$ lower than what we 
found in NGC\,4441 (N(H\ {\sc i}) $\rm = 1.1\cdot 10^{21}\,cm^{-2}$), so no star 
formation occurs in the regular early-type
galaxies of the \citet{2006MNRAS.371..157M} sample. However, considering the
likely S+E merger origin of NGC\,4441, it is possible that the elliptical
progenitor possessed such a gas disc, which would explain why NGC\,4441 has
two \hi tidal tails and otherwise resembles an S+E merger remnant.\\
The low column density \hi disc around ellipticals might also explain why
the star formation in NGC\,4441 is not as enhanced by the merger as in 'normal'
disc--disc mergers.  

\subsection{The formation of the central \hi disc}
Clearly, NGC\,4441 is in an advanced merger stage, because even the progenitor 
nuclei are completely merged in addition to the pronounced stellar tidal features.
In this stage a stable gas disc can already have formed, as shown by 
\cite{2002MNRAS.333..481B}. In unequal-mass mergers the disc formation is more 
impressive \citep{2001hsa..conf...85B}, which again supports that NGC\,4441 is at 
least not an 
advanced remnant
of a merger between two large disc galaxies.   \\
The fairly regular rotation pattern in the central HI disc indicates that 
the merger 
event has 
happened several hundred million years ago 
\citep[e.g.,][]{1999Ap&SS.266..195M}. The gas in this advanced merger has already 
started
 to settle down, but without 
reaching an equilibrium state yet.
The lack of \hi in the centre (Figs.\,\ref{n4441chap_n4441hi}, \ref{n4441chap_pv}) 
might be caused by an efficient
transformation of atomic to molecular gas. This is supported by our CO 
detection within the central 22\arcsec \citep{n4441co}. \\
The velocity gradient of the central disc connects to that of the two tidal 
tails, 
even though perturbations are clearly present. That means that the material in the tails
and the central disc share the same sense of rotation. This indicates that 
a large
amount of angular momentum still remains in the gas of the tidal tails. 
A merger geometry in which the 
angular momentum does not cancel out much seems therefore likely, e.g., a 
prograde merger with parallel rotation spins of the galaxies.
Independent of the presence of an \hi disc in the elliptical progenitor, a 
prograde merger is likely in which the rotation of the spiral and the movement of 
the spiral 
towards the elliptical are aligned. \\
The amount of conserved angular momentum in the gas also depends on other 
factors, e.g., the presence of a bar formed during the interaction in which the
gas loses angular momentum by shocks (see above).
\\
If no strong shocks occur, the gas retains much of its angular momentum,
 which
 prevents the gas from falling toward the central region on 
short timescales. Instead, a gaseous disc will be formed. The remaining 
gas in the tidal tails will finally return to the main body and
will lead to the growth of the present \hi disc. This behaviour is modelled 
by e.g.,
\cite{2002MNRAS.333..481B} and is observed in other merger remnants
like NGC\,7252 
\citep{1994AJ....107...67H,1995AJ....110..140H} and NGC\,4038/39 
\citep{2001AJ....122.2969H}.\\

\subsection{The star forming region}
NGC\,4441 has a moderate ongoing star formation rate of $\rm 1-2\,M_{\odot}\ 
yr^{-1}$. However, there are indications based on optical spectroscopy that
there has been a period of enhanced star formation $\sim$ 1\,Gyr ago 
\citep{1981A&A....97..302B,2005AIPC..783..343M}. \\
The extent of the star-forming region can be investigated using the
continuum data.
The deconvolved size of the 20\,cm continuum source was determined as 
$<$1\,kpc
 (Table\,\ref{n4441chap_contflux}). The emission comes likely 
from ongoing star formation only, since the $q$ value agrees with the 
FIR-to-20\,cm continuum ratio typical for star formation, and based on 
optical spectroscopy 
there are no hints for another origin, e.g. an AGN \citep{1981A&A....97..302B}. 
\cite{n4441co} found molecular gas extended out to $\sim$4\,kpc in NGC\,4441.
Thus, the region of ongoing
star formation 
is embedded in the reservoir of raw material for star formation, namely
the molecular gas extent, but does not cover its full extent. Compared to
the Medusa, where the region of 
ongoing star formation is extended on a scale of $\sim$ 2\,kpc 
\citep[e.g.,][]{1990ApJ...364..471A} and much more intense ($SFR_{\rm opt} 
\sim$ 30\,$\rm M_{\odot}\ yr^{-1}$), the star forming
 activity
seems to be shrunk in intensity and maybe also in spatial
extent. 
 The question arises, why the starburst in NGC\,4441 has already
faded, while the Medusa is still intensely forming
stars. \citet{2000A&A...362...42A} estimate that the burst in the Medusa will
continue only for $\rm \sim 40 \cdot 10^6$ years, under the assumption that
the current star formation rate will not change dramatically. If the
star formation rate depends mainly on the duration of the burst and
decreases with ongoing burst age, the phase
of intense star formation in a merger event, which NGC\,4441 
and the Medusa both underwent, might be very short, and the Medusa will fade
soon 
as well. \\
Furthermore, it has become clear by simulations of S+S mergers that 
the starburst intensity in mergers of unequal masses strongly depends on 
the structure
of the progenitors (e.g., the presence of a bulge) and the geometry of the
merger event (orbits, spin orientation of the galaxies etc.)
\citep{2007arXiv0709.3511C}. Even though no detailed simulations
exist to investigate this phenomenon in the case of S+E mergers, it is
likely that these factors play a similar important role in S+E mergers.
Therefore, (unequal-mass) S+E mergers are 
expected to show a much larger scatter in starburst strength even if they are
in the same merger stage.
\\

\section{Summary}

   \begin{enumerate}
     \item We observed the moderate luminosity merger NGC\,4441 with the
     Westerbork radio interferometer and found a complex, extended \hi structure
     affected by tidal forces. The total \hi mass adds up to
     ${\rm 1.46 \cdot 10^9\,M_{\odot}}$.
     \item We detected two \hi tails. Both tails have a similar extent out
     to 4\arcmin (42\,kpc, southern tail) and 4.6\arcmin (48\,kpc, northern tail)
     away from  
     the centre and they emerge from opposite sides of the main galactic
     body.
     \item The northern \hi arm follows the optical tail closely, however the
     \hi is 
     more extended and shifted to the east. In contrast, the southern \hi arm is
     not aligned with the shell 
     structure found in the optical, but touches only the inner optical shell,
     whereas in the outer shell no neutral gas is found.
     \item The velocity field shows a fairly regular rotation pattern in the inner
     region, associated with the optical main body. The overall velocity gradient is continued in the tidal tails. 
     The southern tail has a generally lower velocity as
     the system, the velocity of the northern tail is larger than the central
     velocity.   
     \item From the 20\,cm continuum flux we calculated the star formation
     rate as 2.4\,\msun\ yr$^{-1}$. From the FIR fluxes we get a SFR of
     1.0\,\msun\ yr$^{-1}$. The resulting ratio $q$ is 2.4, which is in good agreement
     with the typical q value in the literature. We determined a
continuum source size of less than 1\,kpc.      
\item Because of the large differences in the stellar and gaseous
  distribution, it is unlikely that NGC\,4441 is a merger between two large
  disc galaxies. If differences are found in those mergers, they are on much
  smaller scales and can be best explained by superwinds 
  \citep[see][]{2000AJ....119.1130H}. In NGC\,4441, no 
  such 
  strong galactic winds are expected, because the galaxy is in a phase of
  moderate star formation. A merger between two ellipticals can be also
  excluded, because we found a large amount of atomic and molecular gas and
  signs of a past strong star formation activity. This properties are not
  expected in E+E mergers. Thus, it is most likely, that NGC\,4441 is the
  remnant of an S+E merger.  
   \end{enumerate}

\begin{acknowledgements}
 The WSRT is operated by the Netherlands Foundation for Research in Astronomy
 (ASTRON), with support of the Netherlands Organisation for Scientific
 Research (NWO). This research has made use of the NASA/IPAC Extragalactic
 Database (NED) which is operated by the Jet Propulsion Laboratory, California
 Institute of Technology, under contract with the National Aeronautics and
 Space Administration. The research was supported by the German
 Science 
 Organisation (DFG) through the Graduiertenkolleg 787.

\end{acknowledgements}

\bibliographystyle{aa}
\bibliography{7584}

\end{document}